\documentclass[preprint,aps,floatfix,groupedaddress,showpacs,showkeys,preprintnumbers,amsmath,amssymb]{revtex4}
\usepackage{graphicx}
\begin{document} 

\title{Nuclear expansion with excitation}

\author{J. N. De$^{1,2}$, S. K. Samaddar$^{1}$, X. Vi\~nas$^{2}$, and
M. Centelles$^{2}$}
\affiliation{
$^1$Saha Institute of Nuclear Physics, 1/AF Bidhannagar, Kolkata
{\sl 700064}, India \\
$^2$Departament d'Estructura i Constituents de la
Mat\`eria, Facultat de F\'{\i}sica, \\
Universitat de Barcelona, 
Diagonal {\sl 647}, {\sl 08028} Barcelona, Spain}


\begin{abstract}
The expansion of an isolated hot spherical nucleus with excitation
energy and its caloric curve are studied in a thermodynamic model with
the SkM$^*$ force as the nuclear effective two-body interaction. The
calculated results are shown to compare well with the recent
experimental data from energetic nuclear collisions. The fluctuations
in temperature and density are also studied. They are seen to build up
very rapidly beyond an excitation energy of $\sim 9$ MeV/u.
Volume-conserving quadrupole deformation in addition to expansion
indicates, however, nuclear disassembly above an excitation energy of
$\sim 4$ MeV/u.
\end{abstract} 



\maketitle

Understanding the density evolution of nuclear systems at moderate
excitation energies ($\sim 2-8$ MeV/u)
is of much
contemporary interest, both from the theoretical and the experimental
point of view. This is particularly relevant in the context of nuclear
multifragmentation at intermediate 
energy heavy ion collisions \cite{bon,gro,cho1}. With excitation, the
thermal pressure pushes the system towards expansion. At sufficiently
high excitations, the system is ultimately driven towards the {\it
break-up} density below which it ceases to exist in a mononuclear
configuration and ultimately disassembles into many fragments. This
has an important bearing in modelling the equation of state \cite{sto}
of hot nuclear systems; it is also of utmost importance in understanding
explosive nucleosynthesis \cite{ish,bot} in the  
astrophysical context.

Experimental studies have been suggestive of decreasing break-up density 
with increasing excitation energy. Break-up densities have been determined
from studies of correlation functions of emitted light particles
from the source \cite{frit}. It is found that this density
in general decreases with increasing excitation energy in the fragmenting
system, however, the method of analysis leaves room for large uncertainties.
QMD transport model calculations indicate that for medium-heavy systems, 
for excitation energy less than $\sim $  7 MeV/u, densities as low as
$\sim$ 0.35$\rho_0$ can be reached \cite{cib} 
 where $\rho_0$
is the ground state density.
Break-up densities have also been determined from
Coulomb barriers required to fit the intermediate mass ejectile
spectra \cite{bra,vio}; they have as well been determined from the 
analysis of apparent level density parameters required to fit 
the measured caloric curves \cite{nat}. At an excitation energy
of $\sim 8$ MeV/u, the deduced density comes down to $\sim
0.3-0.4 \, \rho_0$, 
but in the excitation energy range explored, the last two sets of 
results are not exactly in consonance; break-up densities derived
from Coulomb barrier systematics are lower
at higher excitations compared to those derived from caloric curve data. 

 From a theoretical standpoint, statistical models 
\cite{bon,gro,cho1}  
have been quite successful in explaining various
observables related to nuclear fragmentation.
The key parameter in these models is the 
{\it freeze-out} density $\rho_f$. On reaching $\rho_f$, the hot
nucleus
undergoes one-step prompt multifragmentation and the interaction
among the generated fragments is assumed to be frozen out
to change the fragmentation pattern. 
The freeze-out density is generally taken to be independent 
of the excitation energy. 
 This density
differs appreciably in different models.  
 Whereas in the canonical or microcanonical models
proposed in Refs.~\cite{bon,gro}, the density $\rho_f$ is
$\sim 0.12-0.2 \, \rho_0$, the corresponding density
in the lattice-gas model \cite{das,cho} is $\sim 0.3-0.4 \, \rho_0$.
Such an uncertainty in both the theoretical and the experimental arena
warrants a closer look at the relationship of the nuclear density
with excitation energy. To explore the break-up
density of a hot metastable mononuclear configuration,
 Sobotka {\it et al} \cite{sob}
have recently performed a calculation where they utilise the fact 
that an isolated system with a given excitation energy pursues 
the maximal entropy configuration for equilibrium. This calculation
is schematic and  involves a few {\it ad hoc} parameters;  the results
compare favorably with the data reported in Ref.~\cite{nat}.
In the present communication, we address this problem with some basic
microscopic inputs starting with a Skyrme type effective two-body
interaction.
We use the celebrated SkM* force \cite{bra1}, well known
for its success in the description of a variety of ground-state
properties of nuclei and diverse phenomena like nuclear
fission and nuclear collective modes.

The experimental perspective in the laboratory is recalled briefly.
When two nuclei collide at intermediate energy, a hot nuclear system
(which may be initially somewhat compressed; the possible
resulting collective flow is ignored in this work)
is formed which may be described statistically by an
effective temperature $T$. If the system were in a heat bath 
at the temperature $T$ (canonical ensemble), the system
could be described by equilibrium thermodynamics driving it to the 
minimum of the free energy. However, the system 
as prepared, is isolated with a
fixed total excitation energy (microcanonical ensemble).
The unbalanced thermal pressure induces
expansion of the system in search of maximal entropy where the total
pressure vanishes and the system is in equilibrium in a mononuclear
configuration. The energy of expansion is derived from the thermal
energy, the temperature thereby decreases. The density at this
maximal entropy state is the lower limit (break-up density) 
for a mononuclear configuration with a fixed excitation energy. 
The system may, however,
gain further entropy from nuclear disassembly.
A full-fledged study of the nuclear disassembly path is
not taken up here, it is mimicked through a volume conserving
deformation of the system at the various stages of expansion.
For simplicity, we consider only quadrupole deformation.

The initial state of the system is prepared by subjecting it in a heat
bath at a chosen temperature $T$. Employing the SkM$^*$ interaction
\cite{bra1}, this is done in a finite temperature Thomas-Fermi
framework within the subtraction scheme \cite{bonc,sur}, well suited
for the description of hot nuclei. The system so prepared
is then detached from the heat bath and allowed to expand with
constant total energy in pursuit of the maximal entropy state. The
expansion is simulated through a self-similar scaling approximation
for the density:
\begin{eqnarray}
\rho_{\lambda }(r)=\lambda^3\rho (\lambda r) \,,
\end{eqnarray}
where the scaling parameter $\lambda$ lies in the range $0< \lambda
\le 1$ and $\rho (r)$ is the base density profile generated in the
Thomas-Fermi procedure.

In actual calculations, we first fix the total excitation energy
$E^*$. The base density profile of the system is generated at a chosen
temperature $T$ such that the excitation energy for this density profile
is less than the given $E^*$. The system is then
allowed to undergo a self-similar expansion till the total excitation
energy (thermal plus expansion) reaches $E^*$ at some value of
$\lambda < 1$. For any density profile, the excitation energy is
calculated as
\begin{eqnarray}
E^*= E(\lambda ,T)- E(\lambda =1,T=0)
\end{eqnarray}
from the Skyrme energy density functional \cite{bra1}.
The corresponding entropy for the expanded configuration $S(\lambda ,T)$ 
is computed. The calculations are repeated for different $T$;
the configuration corresponding to the maximum of 
the entropy profile $S(\lambda ,T)$ 
is the desired equilibrium configuration at the energy $E^*$.


In the subtraction procedure \cite{bonc,sur} the base density for the
hot nucleus is given by $\rho(r) = \rho_{ng}(r)-\rho_g(r)$, where
$\rho_g(r)$ is the density of the surrounding gas representing evaporated
nucleons in which the system is immersed to maintain equilibrium 
at the temperature $T$
and $\rho_{ng}(r)$ is the density of the nucleus-plus-gas
system. The density profile $\rho(r)$ of the heated nucleus is then
independent of the size of the box in which calculations
are done, the density and pressure being zero at large distances.
In the Thomas-Fermi method, the densities $\rho_i(r)$ for neutrons
or protons ($i$ stands for $ng$ or $g$) are given by
\begin{eqnarray}
\rho_i(r)=\frac {1}{2\pi^2} \! \left[\frac
{2m^*_i(r)}{\hbar^2}\right ]^{\frac {3}{2}} \!\!
\int_{V_i(r)}^{\infty} \! \sqrt {\varepsilon -V_i(r)}
f(\varepsilon ,\mu ,T ) \, d\varepsilon .
\end{eqnarray}
Here $m_i^*(r)$ is the  effective $k$-mass
of the nucleon, $V_i(r)$ is
the single-particle potential, $f$ is the Fermi occupation factor,
and $\mu$ is the chemical potential which is same in both
the $ng$ and $g$ phases. The effective mass, single-particle
potential, and the chemical potential are isospin dependent.
The chemical potentials are determined
from particle number conservation ($N$ is the neutron or proton
number): 
\begin{eqnarray}
N= \int g(\varepsilon ,T) f(\varepsilon ,\mu ,T)d\varepsilon ,
\end{eqnarray}
where the single-particle level density $g(\varepsilon ,T)$ 
of the hot nucleus 
in the subtraction procedure is given by \cite{sur}
\begin{eqnarray}
g(\varepsilon ,T) & = &\frac {4\sqrt 2}{\pi \hbar^3}\int
\left [(m^*_{ng})^{\frac {3}{2}}\sqrt {\varepsilon -V_{ng}(r)} 
- (m_g^*)^{\frac {3}{2}}\sqrt {\varepsilon -V_g(r)}\right ]r^2 dr.
\end{eqnarray}
The effective mass, single-particle
potential, and the chemical potential 
are evaluated at the appropriate temperature and 
scaled densities, so also the
single-particle level density and the entropy.
 The latter reads as
\begin{eqnarray}
S(\lambda ,T)=-\int g(\varepsilon ,T)\left 
[f \ln f + (1-f) \ln (1-f)\right ] d\varepsilon .
\label{entrop}
\end{eqnarray}
The total entropy is the sum of the neutron and proton contributions.

For our study, we have chosen $^{150}$Sm as a representative system.
The central density $\rho_c$ of the
calculated mononuclear equilibrium configuration
in units of $\rho_0$ (the central density of the unexpanded nucleus
at $T=0$) \cite{fn} 
is displayed in Fig.~1 as a function of the excitation energy and
compared with the experimental data. The thin (dashed and solid) lines
correspond to the canonical ensemble calculations,
{\it i.e.}, when the system is in a heat bath; 
the thick lines are those for the isolated expanded nucleus at
equilibrium. The filled circles are the experimental points extracted
from the apparent level density parameters \cite{nat} for the mass
selection $140<A<180$, where $A$ is the mass number of the system, and
the empty squares are the ones obtained from Coulomb barrier
systematics \cite{bra,vio} for Au-like systems.

The effective mass $m^*$ defined previously comes from the momentum
dependence of the single-particle potential, which is the $k$-mass
$m_k$. However, $m^*$ should have a frequency
dependent mass-factor $m_\omega /m$:
\begin{eqnarray}
m^*=m \frac {m_k}{m} \frac {m_{\omega }}{m} .
\end{eqnarray}
The $\omega$-mass originates from the coupling of the single-particle
motion with the collective degrees of freedom. This has the effect of
bringing down the excited states from high energy to lower energy near
the Fermi surface, thus increasing the many-body density of states 
[$m_{\omega}/m \ge 1$, see Eq.~(8)] at low excitations.
It may {\em a priori} have a significant role to play in the present
context as the system can accommodate comparatively more entropy at a
given excitation energy.
The self-consistent evaluation of the $\omega$-mass is beyond
the scope of the present work; we use the phenomenological
form \cite{pra,shl}
\begin{eqnarray}
\frac {m_{\omega}}{m}= 1 - 0.4 \, A^{\frac {1}{3}} \exp\left
[-\left (\frac {T}{21A^{-\frac {1}{3}}}\right )^2\right ]
\frac{1}{\rho(0)} \frac{d\rho (r)}{dr} ,
\end{eqnarray}
where $\rho (r)$ is the density profile at temperature $T$. The effect
of $m_{\omega}$ is incorporated \cite{pra,de} by replacing the
single-particle potential $V$ with $(m / m_{\omega}) V$ in
Eqs.~(3)--(5).

In Fig.~1, the solid (dashed) lines refer to calculations performed
with (without) inclusion of $m_\omega /m$.
With increasing temperature, $m_{\omega}/m$ 
tends to unity.
We find that the values of $\rho_c$ calculated at moderate or higher
excitations with and without the inclusion of the $\omega$-mass
are practically the same; however, the relative densities
$\rho_c /\rho_0$ for the
two cases run parallel there because the 
values of $\rho_0$ are different in the two situations.
The calculations for the canonical ensemble (thin lines)
terminate at an excitation energy of $\sim 5.5$ MeV/u
corresponding
to $T \simeq 8.5$ MeV, the limiting temperature
\cite{bonc,sur} for this nucleus beyond which it is unstable
in a heat bath. The thick curves, as compared to the thin ones,
show appreciably lower values of density that compare 
reasonably with the  densities extracted from the analysis of 
the caloric curve measurements performed in \cite{nat} (filled circles).
The importance of a microcanonical
treatment is thereby indicated. The fit with the data obtained 
from Coulomb barrier systematics (filled squares)
\cite{vio}  is relatively poor; however,
an ambiguity in their extraction procedure
has been pointed out recently \cite{rad}.

The $\omega$-mass does not appear to have a very distinctive role 
to the density evolution with excitation energy. In the excitation
range indicated in the figure, it turns out that with inclusion
of the $\omega$-mass, the equilibrium configuration corresponds to a 
lower value of the temperature as compared to that obtained with
$m_\omega /m$=1; this tends to increase the central density
$\rho_c$ in the former case.
 On the other hand, the scale parameter $\lambda$
is found to be comparatively lower with inclusion of the 
$\omega$-mass, resulting in a lower value of $\rho_c$.
The combined effect of these opposing tendencies then results in
the near equality of $\rho_c$ in both the calculations.

The expansion of the nucleus
has an important bearing on the correlation of the excitation
energy with temperature.
The caloric curve so obtained for the expanded hot nucleus $^{150}$Sm
in equilibrium is displayed in the upper panel of Fig.~2.
 At a fixed excitation energy, the system
cools down with expansion and therefore the recorded temperature
at the equilibrium configuration is significantly lower than that for
the unexpanded nucleus prepared initially with the
same excitation. In the figure, $T$ refers to the canonical
temperature. We have checked that the microcanonical
temperature obtained from $T^{-1} = \partial S_{\rm eq}/ \partial E^*$, 
where $S_{\rm eq}$ is the total entropy and $E^*$ the excitation
energy at equilibrium, is not much different from the canonical one.
For comparison, a representative set
of experimental data for medium mass nuclei
\cite{cib} are also shown in the figure. The notations used have the
same meaning as in Fig.~1. The effect of isolation is not appreciable
at relatively lower excitations, but is more manifest at higher
excitations. The calculated microcanonical results are seen to agree
nicely with the experimental caloric curve. The canonical caloric
curves with and without the inclusion of the $\omega$-mass merge at
high temperatures because $m_\omega /m$ does not differ much
from unity. The microcanonical caloric curves, on the other hand,
even at high excitations do not merge; the highest temperature
encountered in these calculations is $\sim 8$ MeV where
$m_\omega /m$ is typically $\sim 1.02$ at the surface  region
affecting the equilibrium configuration a little which is amplified in
the caloric curve.

At a given temperature, the system is found to have more excitation
with inclusion of the $\omega$-mass as more states are available near
the Fermi surface to absorb more energy. Beyond $E^*/A \sim 6$ MeV,
$T$ seems to saturate and at $E^*/A \sim 8$ MeV, a downward slope in
the caloric curve is apparent for the isolated system implying
negative heat capacity. The expansion energy [$E_{\rm
expn}=E(\lambda ,T)-E(\lambda =1 ,T)$] comprises a significant part of
the total excitation at higher values of $E^*$ inducing the
above-mentioned characteristics in the caloric curve. The expansion
energy is found to be in good agreement with that obtained in the
Expanding Emitting Source model of Friedman \cite{fri} as shown in the
lower panel of Fig.~2. The filled diamond refers to an experimental
estimate \cite{vio}, which is close to the prediction from our
calculations.

The entropy profile obtained at a fixed excitation energy may be
employed to calculate the mean values and the fluctuations 
around the mean of the temperature and of the density or volume.
The probability of finding a configuration with scale parameter
$\lambda $ and temperature $T$ at a fixed excitation energy
$E^*/A$ is given by 
\begin{eqnarray}
W(\lambda,T) \propto e^{S(\lambda ,T)} ~,
\end{eqnarray}
where $S(\lambda ,T)$ is the total entropy (\ref{entrop})
 of the given configuration.
The $n$-th moment of the central density $\rho_c$ is then given by 
\begin{eqnarray}
\langle \rho_c^n(\lambda ,T) \rangle = \frac {\int 
e^{S(\lambda ,T)} \rho_c^n(\lambda ,T) d\rho}
{\int e^{S(\lambda ,T)} d\rho} ~,
\end{eqnarray}
which allows to calculate the mean $\langle \rho_c \rangle$ and the 
variance $\sigma_{\rho}^2= \langle \rho_c^2 \rangle 
-\langle \rho_c \rangle ^2$. Similarly, the mean and variance
of the temperature at a fixed excitation can be evaluated.

For a thermodynamic system, the average and the most probable
(equilibrium) value of an observable are the same. For a finite
system, however, they may be different. Experimentally, the
average value is the relevant quantity.
The average and the most probable values 
of the temperature and the specific volume $v_c$ (= 1/$\rho_c$)
in units of $v_0$ (= 1/$\rho_0$) for the system
considered are displayed in the upper
and lower panels of Fig.~3. It is seen that the differences between
the averages and the most probable values of the temperature and the 
break-up volume (or density) are not very significant.
The fluctuations $\sigma ^2$ in
temperature and the specific volume (measured in units of $v_0$)
are shown in Fig.~4. The fluctuations rise smoothly up to the
excitation energy $E^*/A \sim$ 9 MeV, beyond which this 
build-up is very sudden; this is particularly more pronounced 
for the volume fluctuations. This large density fluctuation indicates
that beyond $E^*/A \sim$ 9 MeV the system becomes unstable and
breaks up in many pieces. It turns out that the negative branch
of the specific heat (Fig.~3) and the large fluctuations start
at around the same excitation energy. A possible close correlation
between them is thereby indicated.

So far, for the search of the maximum entropy configuration, the shape of
the excited expanding system has been constrained to a spherical one;
a possible deformation path along with expansion might also contribute
additional entropy and mimics a fragmentation channel. To investigate this
aspect, at all stages of the expansion at a fixed excitation energy,
a volume conserving quadrupole deformation is explored.
In a volume conserving deformation, only the surface and Coulomb
energies change. To calculate these changes, a sharp surface approximation
to the density profile is made ($R_{\rm sharp}=\sqrt{\frac {5}{3}
\langle r^2
\rangle}$) which also facilitates the calculation of entropy from deformation.
At a deformation $\beta$, the excess Coulomb energy of the expanded
deformed nucleus is $\delta E_c(\lambda ,\beta )=E_c(\lambda ,0)f(\beta )$;
the function $f(\beta )$ is given in Ref.\cite{eis}. The surface free energy
due to deformation is given by
$\delta F(\lambda ,\beta )=\delta {\cal A}(\lambda ,\beta )\sigma (\rho ,T)$,
where $\delta {\cal A}$ is the excess surface area of the nucleus arising out
of deformation. The surface tension coefficient $\sigma$ at a density $\rho$
and temperature $T$ is taken as 
$\sigma (\rho ,T)=\alpha (T)g(\rho )$.
The temperature dependence \cite{bon1} of the surface tension is given by
$\alpha (T)$. The surface tension has its maximum value at the ground-state
density; for an expanded system, $\sigma$ decreases which can
be well represented by $g(\rho )$ taken to be a polynomial in $\rho $. We
have calculated this density dependence from the prescription of Myers
and Swiatecki \cite{mye} using the scaling approximation to the
ground-state density profile of semi-infinite nuclear matter.
The excess entropy from deformation is then calculated as 
$\delta S= - \partial (\delta F) / \partial T |_\rho$
which immediately gives the excess surface energy
$\delta E_{\rm surf} = \delta F+T \delta S$.
The total excitation energy of the expanded deformed system is then
${\cal E}^*(\lambda ,\beta )= {\cal E}^*(\lambda ,0)+\delta
E_c(\lambda ,\beta )+\delta E_{\rm surf}(\lambda ,\beta )$
and the corresponding entropy is
$S(\lambda ,\beta )=S(\lambda ,0)+\delta S(\lambda ,\beta )$.

Along the deformation path, a barrier is faced coming from the 
interplay of the Coulomb and surface energies; it decreases with
increasing temperature and increasing expansion. In actual calculations,
the scale parameter $\lambda$ is adjusted such that for a chosen
temperature, the excitation energy at the top of the barrier
 ${\cal E}^*(\lambda ,\beta )$ matches the given excitation $E^*/A$.
This is repeated for different temperatures and the maximum entropy
among these different configurations is selected. If this entropy 
exceeds that for the expanded spherical equilibrium
configuration, then the deformed shape is favoured leading to the
fragmentation channel. 

This extra entropy gained due to deformation 
(we call this $\Delta S$) over that at the spherical
equilibrium shape is shown in the top panel of Fig.~5 as a function of 
excitation energy. For $E^*/A$ less than $\sim$ 4 MeV, $\Delta S$ is 
negative in our chosen restricted deformation space and so a spherical
equilibrium configuration is more probable. Above this
excitation energy, fragmentation resulting from deformation is
more favourable. Beyond $E^*/A \sim$ 9 MeV, the barrier vanishes 
and the system undergoes spontaneous fragmentation from the spherical
equilibrium configuration.

In the middle panel of Fig.~5, the equilibrium central densities with (solid line)
and without (dashed line) deformation are displayed as a function of
$E^*/A$. With deformation, the maximal entropy configuration occurs
at a lower temperature with a smaller scale parameter $\lambda $
resulting in a somewhat reduced density as seen. The corresponding caloric
curves are shown in the bottom panel. All the above calculations pertain to
prolate deformation, oblate shapes are found to have lesser entropy.

We have addressed to some gross features of a mononuclear
configuration at medium and high excitations in
a semi-microscopic framework based upon a realistic effective nuclear
interaction.
The calculated break-up densities for the nucleus in the
microcanonical formulation are in
general in good agreement with the ones extracted from the 
experimental data analysis \cite{nat}.
The generated mononuclear caloric curve
also compares very well with the experimental results. The plateau observed
in experimental data has been taken to be suggestive of a possible
phase coexistence \cite{poc}. 
In our
calculations, the plateau comes naturally from nuclear expansion with
excitation. 
For the bloated spherical mononuclear configuration, the rapid
build-up of fluctuations, particularly in the density, is suggestive 
of the instability of this configuration against prompt multifragmentation
beyond $E^*/A \sim$ 9 MeV. Expansion with deformation degrees of freedom
may have significant effects on the physical observables; to have an orientation
on the role of deformation, we have considered volume-conserving quadrupole
deformation. It is found that above $E^*/A \sim$ 4 MeV the system
favours deformation, a precursor to fragmentation. 
 The global trends provided by the present 
model, particularly with the
inclusion of the frequency dependence in the effective mass, are 
qualitatively consistent with the experimental data
\cite{bra,vio,nat} as well 
as with the ones obtained from 
 other mononuclear formulations \cite{sob,sob1}.
At high excitations, the collective flow may play a significant role
in the nuclear collision process. This has been ignored in the
present calculation; it would be worthwhile to investigate its
effect on the break-up density and also on the caloric curve.

\mbox{ }

J.N.D. and S.K.S. acknowledge the financial support from
DST and CSIR, Government of India, respectively. 
J.N.D. gratefully acknowledges the hospitality
at the University of Barcelona and a financial grant 2003PIVB00077
from Generalitat de Catalunya. 
M.C. and X.V. acknowledge financial support from Grants No.\
FIS2005-03142 from MEC (Spain) and FEDER, and No.\ 2005SGR-00343 from
Generalitat de Catalunya.

\newpage
\centerline
{\bf Figure Captions}
\begin{itemize}
\item[Fig.\ 1]  The calculated equilibrium 
density in units of $\rho_0$ as
a function of the excitation energy per nucleon is compared with the
experimental data. For notations, see text.
\item[Fig.\ 2] The mononuclear caloric curve (upper panel)
and the expansion energy per nucleon (lower panel).
\item[Fig.\ 3] The average and the equilibrium values of temperature
(upper panel) and the specific volume (lower panel) of the nucleus
$^{150}$Sm as a function of excitation energy.
\item[Fig.\ 4] The variance in the temperature (upper panel) and
the specific volume (lower panel) as a function of excitation
energy.
\item[Fig.\ 5] The deformation entropy $\Delta S$ (top
panel), the equilibrium density in units of $\rho_0$ (middle panel)
and the caloric curve (bottom panel). The solid lines represent
the results with deformation and the dashed lines correspond to the
spherical configuration. The experimental points (filled circles
and open squares) are the same as those given in Figs.~1 and 2. 
\end{itemize}

\begin{figure}
\includegraphics[width=0.85\textwidth,angle=0, clip=false]{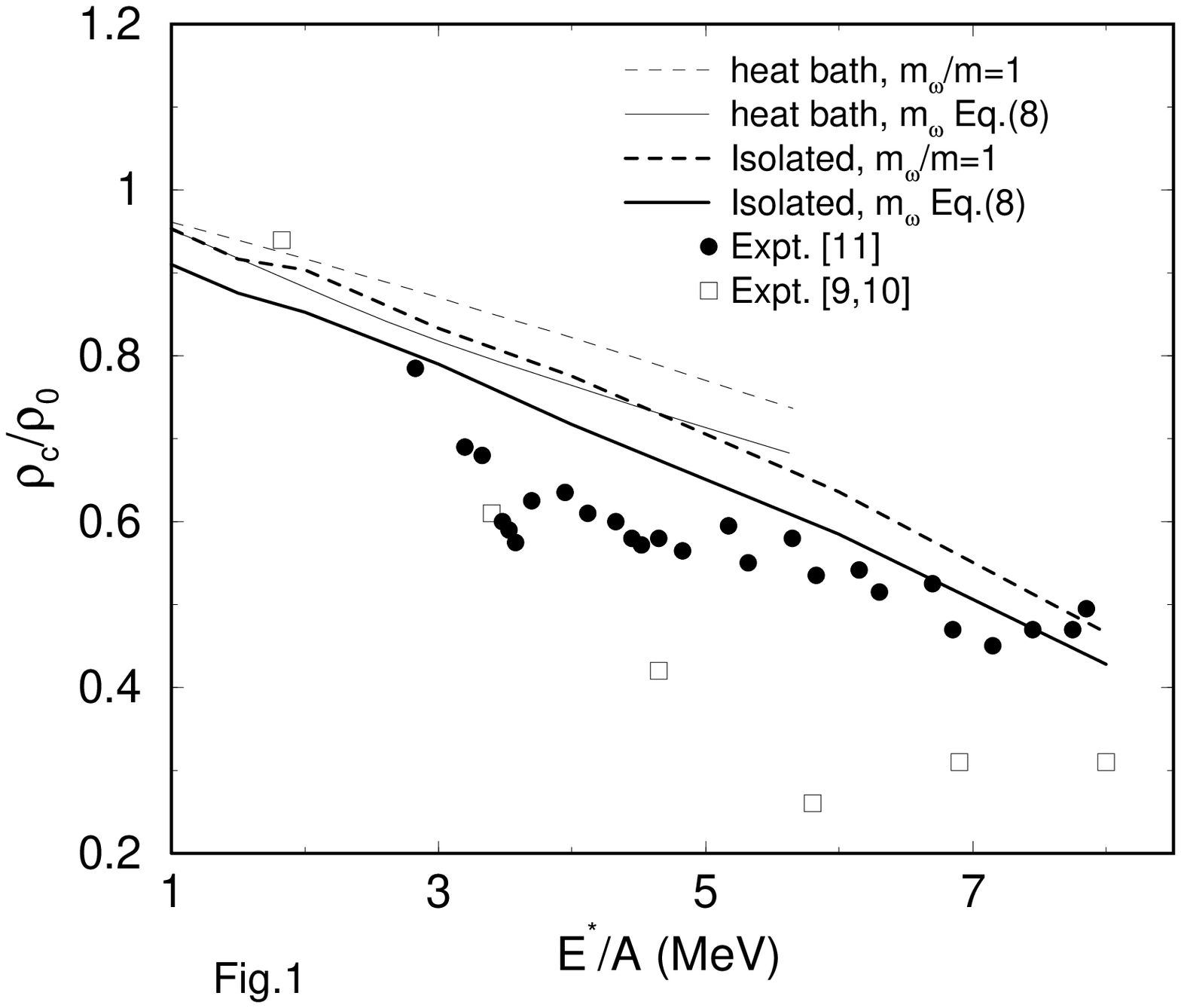}
\end{figure}
\begin{figure}
\includegraphics[width=0.75\textwidth,angle=270, clip=false]{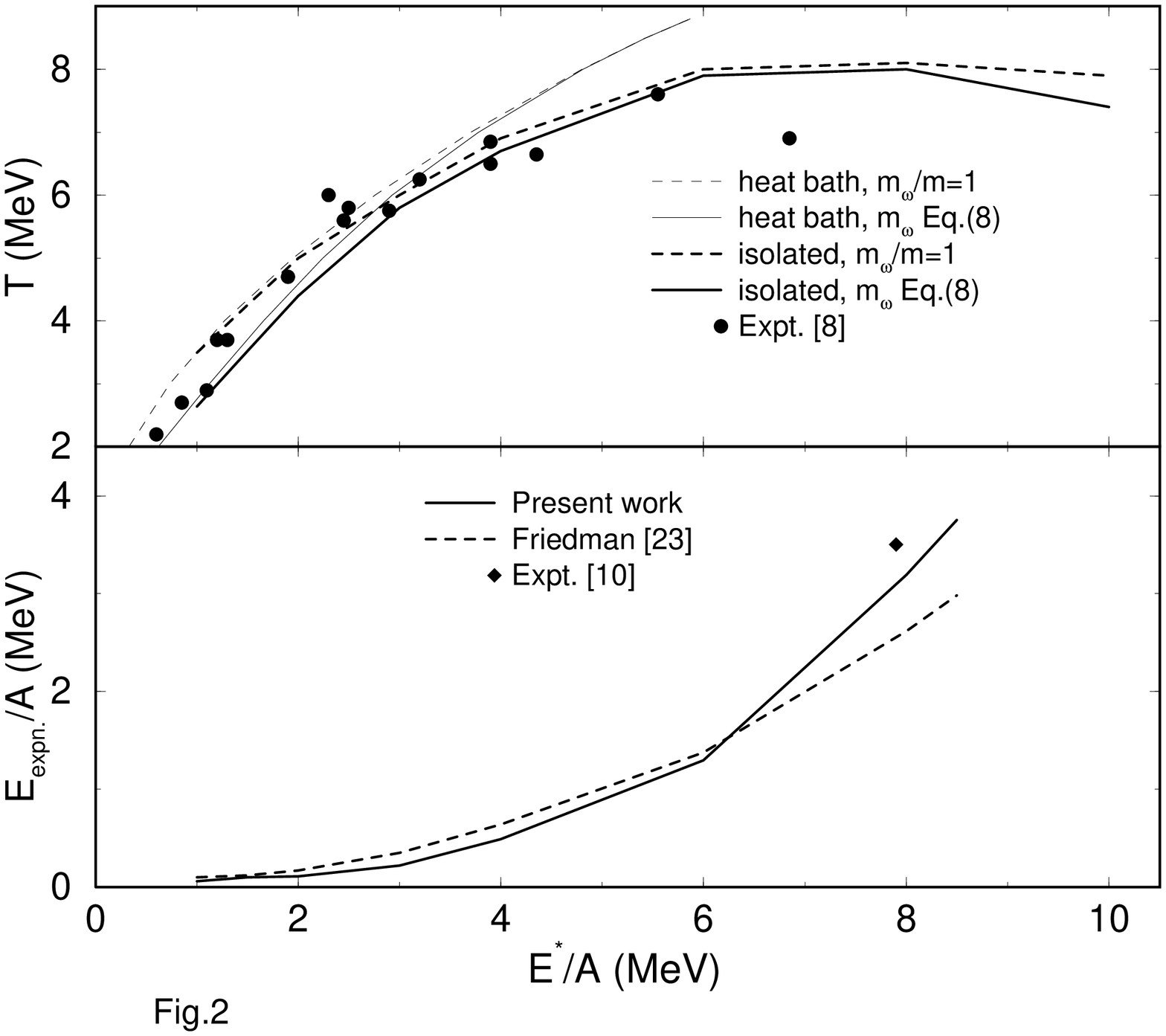}
\end{figure}
\begin{figure}
\includegraphics[width=0.88\textwidth,angle=0, clip=false]{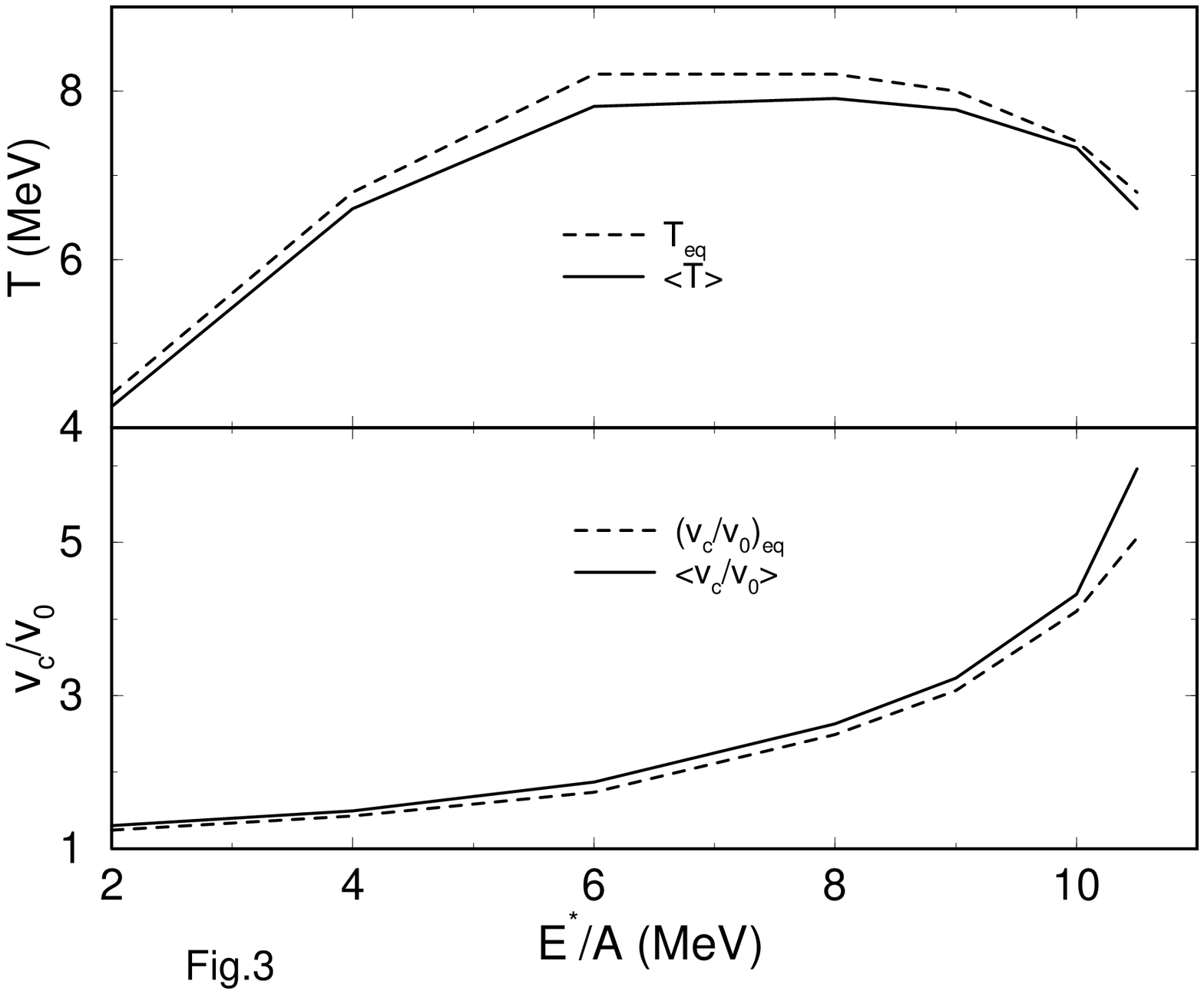}
\end{figure}
\begin{figure}
\includegraphics[width=0.88\textwidth,angle=0, clip=false]{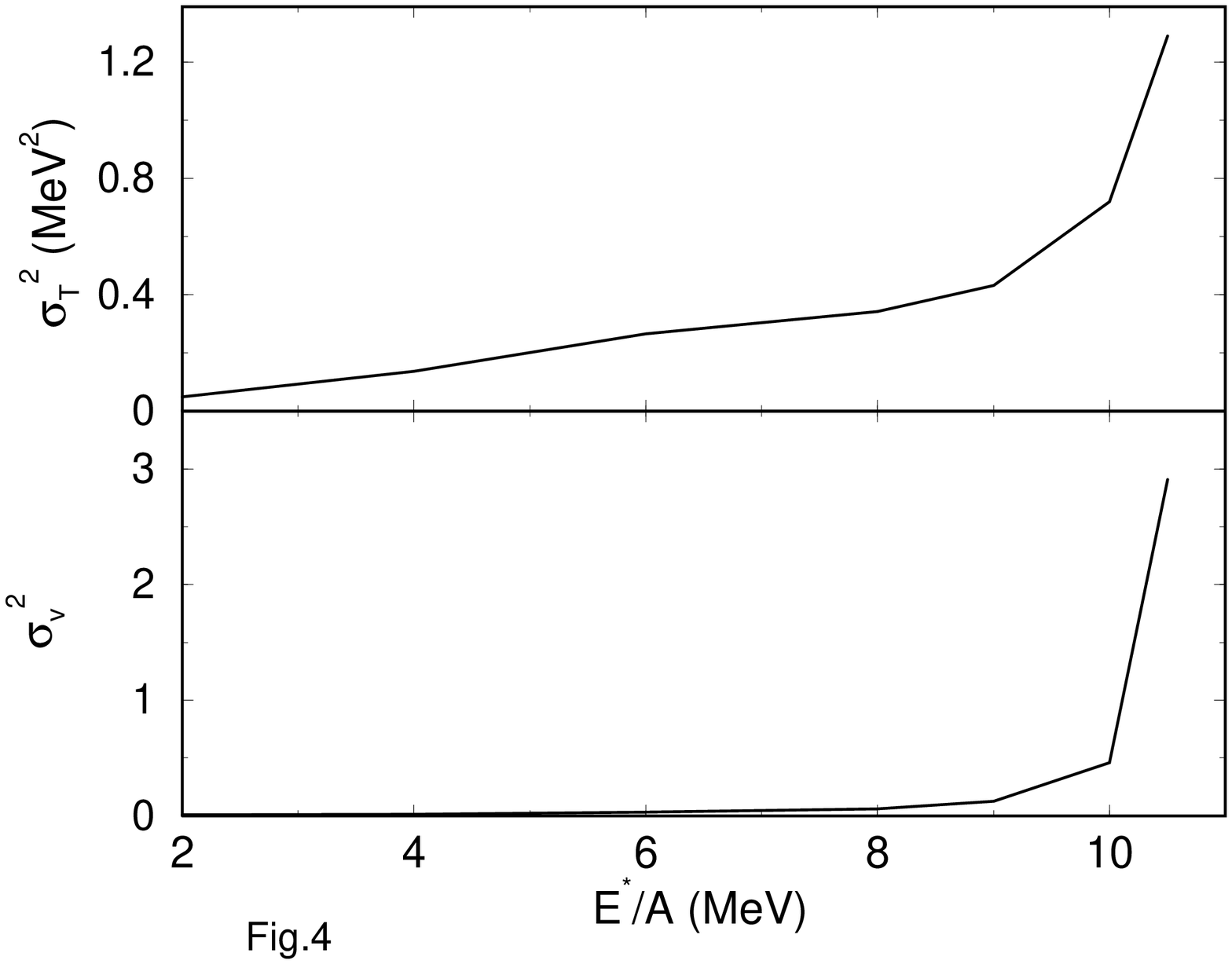}
\end{figure}
\begin{figure}
\includegraphics[width=0.70\textwidth,angle=0, clip=false]{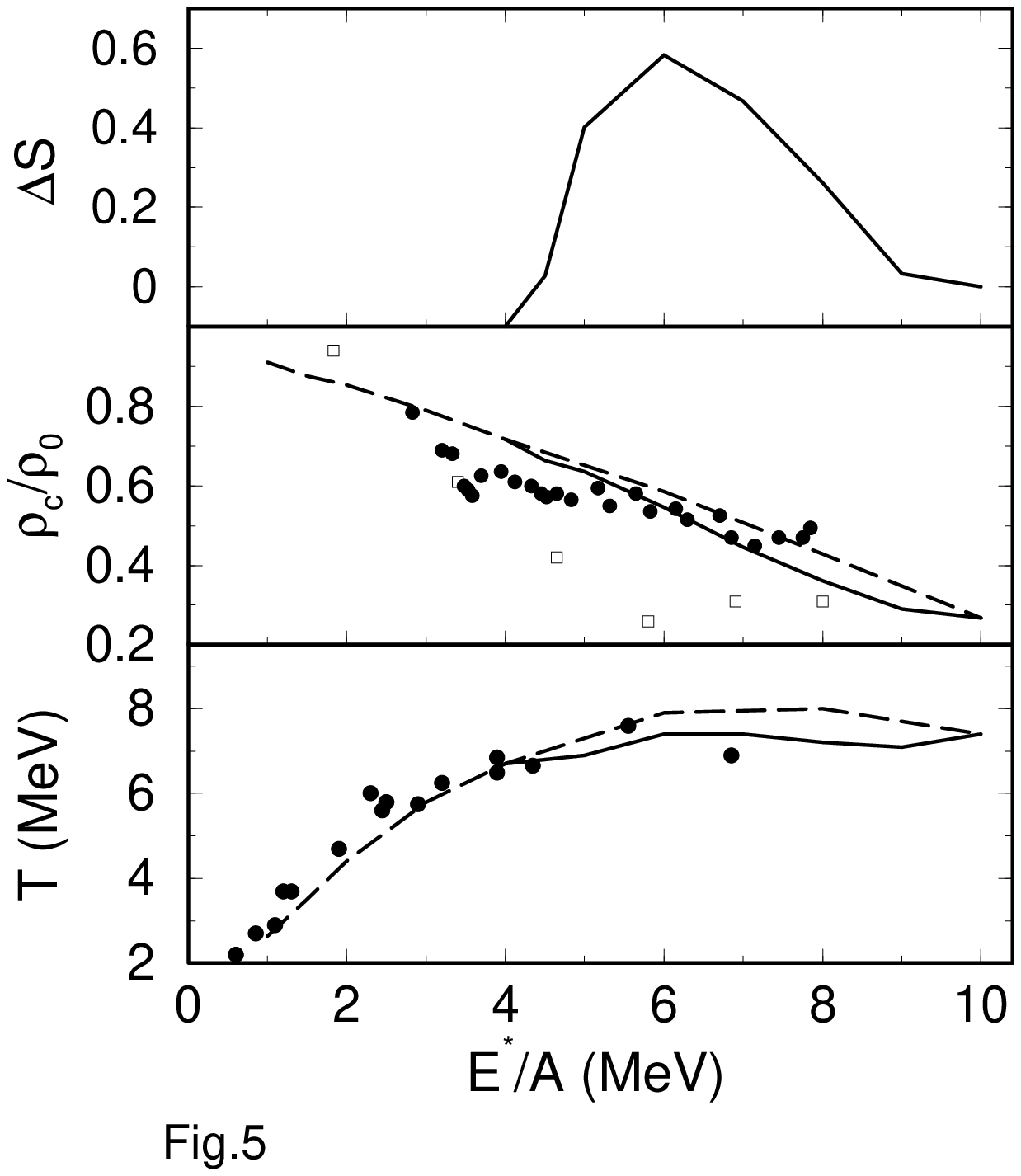}
\end{figure}


\begin{thebibliography}{99}
\bibitem{fn} Since the density has a profile, the use of a single
value for $\rho$ 
has some room for ambiguity. To keep $\rho_c$ and
$\rho_0$ on the same footing, we have taken for $\rho_0$ the value of
the ground-state central density. It has, however, been checked that
the results are nearly insensitive to the choice of
$\rho$ as the corresponding equivalent uniform 
density at  the relevant temperatures.
\end{thebibliography}
\end{document}